\documentstyle[epsf,twoside,fleqn,espcrc2]{article}

\title{Semileptonic Form Factors}

\author{T. Onogi\address{Department of Physics,
        Hiroshima University, \\ 
        1-3-1 Kagamiyama, Higashi-Hiroshima, Hiroshima 739, Japan}}%

\begin{document}
\begin{abstract}
I report the current status of the heavy-light decay constants, 
the bag parameters and the semileptonic form factors.
I compare the heavy-light decay constants with Wilson-Wilson 
and clover-clover fermions. Systematic errors such as
scale setting and renormalization factors are also discussed.
$1/M$ dependences for the heavy-light semi-leptonic
form factors near $ q^2 = q^2_max $ with clover-clover and
NRQCD-Wilson fermions found to be small.
\end{abstract}

\maketitle

\section{Introduction}
 Calculations of 
 the $B$ meson decay constants, the bag parameters and the semileptonic
 form factors have been attracting the attention of lattice
 community over the past several years, since those are
 indispensable ingredients for the determination of $V_{CKM}$
 and then the test of the standard model as well as the study
 of CP violation and the quark mass generation.
 I review the update in Lattice QCD made in this year.
 
 For the formalism of the heavy qaurk on the lattice the
 Fermilab group~\cite{El-Khadra_Kronfeld_Mackenzie_97} 
proposed a systematic method to treat the
 error associated with the large heavy quark mass in the
 relativistic formalism.
 Then the large scale calculations of Fermilab, JLQCD and
 MILC collaborations at several $\beta$ values using the
 Wilson or clover action are giving a reliable prediction for 
 $f_B$. 
 There has also been a considerable development in the
 calculation of $f_B$ using the NRQCD formalism, which was
 reviewed by Ali-Khan in this
 conference\cite{Review_NRQCD_97}. 
 
 New calculations of the semi-leptonic decay form factors
 $B\rightarrow\pi/\rho$ have become available by Hiroshima
 and JLQCD collaborations using the NRQCD and the clover
 actions for heavy quark.
 Their simulations treat the $b$-quark directly without
 extrapolating from the charm quark mass region, and the
 heavy quark mass dependence of the form factors are clarified. 
\section{Heavy Quark on the Lattice}
In currently accessible simulation, the bottom and charm quark masses
in lattice unit are large:  $m_b a = 1 \sim 4$, $m_c a = 0.4 \sim 1$.
Since the brute force extrapolation to the continuum limit is
 not realistic, one has to formulate an effective theory in
which there is no large systematic uncertainty of $O(m_Qa)$.
For Wilson and clover action, Fermilab interpretation is
such an example\cite{El-Khadra_Kronfeld_Mackenzie_97}.
In their formalism the relativistic heavy quark is
interpreted as an effective action in the heavy quark rest
frame, which is only valid in the small spatial momentum regime. 
 
In order for high precision
calculation of the matrix elements, it is important to understand 
 the source of systematic errors.
 In the following, I discuss the expected size of the
 systematic errors in the calculations of the matrix elements
 of the heavy-light mesons with the Wilson/clover fermions.

Besides quenching errors, some of the largest sources of errors 
are perturbative and the lattice discretization errors from the action
and the operators. 

First let us discuss the light quark action.
 Since the lattice cutoff $a^{-1}$ and the typical momentum
 scale of the quark (and gluons) $p$ are the only
 dimensionful quantity,  
 the above errors can be expanded in powers of $ap$ and
 $\alpha_s$.  
As is well known, the leading error for the Wilson light quark 
comes from the lack of the clover term, which is $O(ap)$.
On the other hand, the leading errors for the clover light quark 
 are $O((ap)^2)$, $O(\alpha_s^2)$, and $O(\alpha_s ap)$. 
There exisits heavy-light simulations with 
$c_{sw} =1/u_0^3(1+0.199 \alpha_s)$ \cite{Luscher_Weisz_96} at one-loop 
level by JLQCD, in which case the leading errors of the
action are $O((ap)^2)$  and $O(\alpha_s^2 ap)$ only. 

The situation is similar for heavy quark action.
The errors can again be expanded in powers of 
 $\alpha_s$ and $ap$, except that the coefficients which appear in the 
expansion are now some functions of $a m_0$ where $m_0$ is the 
bare quark mass. In the Fermilab interpretation of the Wilson/clover quark, 
the small momentum expansion of the effective Hamiltonian
reads\cite{El-Khadra_Kronfeld_Mackenzie_97} as,
\begin{eqnarray}
H & = & - ( \frac{1}{2m_2a}+ b^{(1)}\alpha_s + \cdots) \vec{D}^2
\nonumber \\
   &   &  - (\frac{1}{2 m_Ba}+ b_B^{(1)}\alpha_s + \cdots) \cdot i 
 \vec{\Sigma} \cdot \vec{B}
\nonumber \\
  &   & + {\rm ~higher~ order},
\end{eqnarray}
 with $m_2a$, $m_Ba$, $b^{(1)}$, $b_B^{(1)}$,... are
 functions of $m_0a$. 

The tree level coefficients in the above expresssion 
are described as follows for Wilson/clover action,
\begin{eqnarray}
 \frac{1}{2m_2a} & = & \frac{1}{m_0a (2+ m_0 a)}+\frac{1}{1+m_0a} 
\nonumber \\
 \frac{1}{2m_Ba} & = & \frac{1}{m_0a (2+ m_0 a)}+\frac{c_{sw}}{1+m_0a}. 
\end{eqnarray}

The leading error in the Wilson heavy quark action arises in
the spin-magnetic coupling, as $1/2m_Ba$ differs from
$1/2m_2a$ by $1/(1+m_0a)$.
Thus the size of the error is estimated to be of $O(p/m_0)$, 
which may relatively be smaller than the error from the
light quark action.  
For clover heavy quark action, 
the error from the spin-magnetic coupling arises  at one-loop
level, which is $O(\alpha_s ap)$ or $(\alpha_s p/m_0)$. 
Therefore, the leading errors in clover heavy quark are
$O(\alpha_s ap)$, $O(\alpha_s p/m_0)$, $O((a p)^2)$ and
$O((p/m_0)^2)$. 
So far there exists no calculation
of $c_{sw}$ at one-loop for finite quark mass, the knowledge
of which would be useful in removing the error of
$O(\alpha_s ap)$ and $O(\alpha_s p/m_0)$.

Next let us consider the error from the current operator 
by taking the example of the heavy-light axial-vector
current. 
The heavy-light current operator has the following form in
the Fermilab interpretation:
\begin{eqnarray}
A_4 
& = &   
d_0  \overline{\psi_q} \gamma_5 \gamma_4 \psi_Q \nonumber \\
& + &   
d_1  \overline{\psi_q} \gamma_5 \gamma_4 
        \vec{\gamma} \cdot \vec{D}\psi_Q \nonumber \\
& + &   
d_2   \vec{D} \overline{\psi_q} \cdot \vec{\gamma} 
	\gamma_5 \gamma_4 	\psi_Q, \nonumber
\end{eqnarray}
 where
\begin{eqnarray}
d_0 & = &  1 + d_0^{(1)} \alpha_s + \cdots \nonumber \\
d_1 & = &  -\frac{1}{2 m^{\prime}} + d_1^{(1)} \alpha_s + \cdots \nonumber \\
d_2 & = &  d_2^{(1)} \alpha_s  + \cdots \nonumber,
\end{eqnarray}
and
\begin{eqnarray}
-\frac{1}{2m^{\prime}a} & =& -\frac{1}{2m_2 a} 
  + \frac{m_0 a^2}{2(1+m_0a)(2+m_0 a)},
\end{eqnarray}
for both Wilson and clover actions.
The second term in the right hand side of this equation
causes $O(ap)$ error formally, 
but in practice, it is negligibly small because of the small coefficient 
$m_0 a/(2(1+m_0a)(2+m_0 a))$.

Recently, the one-loop correction for the leading term $d_0^{(1)}$ 
was calculated fully incorporating the heavy quark mass
dependence\cite{Ishikawa97}.  
Therefore, for both Wilson and clover actions, 
the leading errors are from the higher derivative terms,
which are of $O(\alpha_s ap)$ and $O(\alpha_s p/m_0)$ in
addition to the $O(\alpha^2)$ and $O((ap)^2)$ errors for
both the Wilson and clover actions. 
For further reductions of the errors, calculation of the
one-loop corrections $d_1^{(1)}$ and $d_2^{(1)}$ is
necessary.  
The NRQCD group performed such calculation for NRQCD heavy
quark\cite{JSHEP}.
They find the effect of operator mixing reduces $f_B$ by
about 10\% at $\beta$=6.0.
It would be interesting to see whether it is also the case
for clover heavy quark case.
 
To summarize, the Wilson heavy-light system has leading
errors of $O(ap)$ and $O(p/m)$, while the clover heavy-light
system has leading errors of $O(\alpha_s^2)$, $O(\alpha_s
ap)$, $O(\alpha_s p/m)$ and $O((pa)^2)$, $O((p/m)^2)$.
Future one-loop renormalization with massive fermion for
both the action and the current remove one of the leading
errors, which is $O(\alpha_s ap)$ and $O(\alpha_s p/m)$. 

\section{Lattice Results}
In this section, we review the numerical simulation results
for the decay constant and the form factors. We discuss whether the 
picture for the systemtic errors in the previous section explains 
the lattice spacing dependence of the actual data or not, which is
of particular importance in extrapolating to the continuum limit.

\subsection{Decay Constants}
Table~\ref{tab:Simulations} shows results on heavy-light 
decay constants by various groups uing Wilson/clover fermions.

MILC collaboration uses Wilson fermion in the Fermilab formalism,
in the quenched approximation
with scales set from $f_{\pi}$ and  $m_{\rho}$ .
They also perform simulations with dynamical configurations
for $\beta=5.445, 5.5, 5.6$.
They get their best preliminary results with the scale from $f_{\pi}$
as shown in Table~\ref{tab:Simulations}.
The systematic erors from quenching are estimated by comparing 
the result with scales set from $m_{\rho}$ and $f_{\pi}$,  
and also the result with $n_f=0$ and $n_f=2$. They also study the
chiral extrapolation error by comparing the results with 
the linear and quadratic fits in the chiral extrapolation.
They use the local axial vector current in KLM normalization,
with $Z_A$ from one-loop perturbation theory in the massless
limit. The renormalization
scale for $\alpha_v$ is $q^{\ast}=2.32/a$\cite{Bernard_Golterman_97}.

JLQCD collaboration uses Wilson/clover fermions in the Fermilab
formalism on three different lattices in the quenched approximation.
They obtain their preliminary results with scales set from 
$f_{\pi}$, $m_{\rho}$ and the string tension with $f_{m_{\rho}}$ for
best results.
They use the local axial vector current in KLM normalization
with $Z_A$ from one-loop pertubation theory, in which the heavy quark 
mass effect is taken into account\cite{Ishikawa97}  
The renormalzation scale is $q^{\ast}=1/a$. 

APE collaboration uses both Wilson/clover fermions 
with $\beta=6.0,6.2$ in the quenched approximation\cite{Allton97}.
The compute the heavy-light decay constants  in the 
charm quark mass range and extrapolates them in
  $1/M$ to bottom quark mass range. 
They first compute the ratio $R \equiv f_{PS}/f_{\pi}$ with 
scales set from $f_{\pi}$, $m_{\rho}$, the string tension
and $f_{K^{\ast}}$, then multply $f_{\pi}^{exp}$. 
They find consistent results for all these four scale settings.
They use local current both in KLM and in standard normalization.
For setting the scale from $f_{\pi}$ using lattice results
they use $Z_A$ determined nonperturbatively using chiral Ward
identities. \cite{Henty95,Martinelli93}

UKQCD collaboration uses clover fermion in the quenched 
approximation.
They use the rotated axial vector current with $Z_A$ from one-loop 
perturbation theory.
They obtain their best results with scales set from $m_{\rho}$
\cite{UKQCD94,Ewing96}. 

Fermilab group performed calculations of the heavy-light decay
constant with clover fermion in Fermilab formalism in the 
quenched approximation. They use the tree-level improved current
with $Z_A$ obtain from one loop perturbation theory in the
massless limit. The scale is set from $f_K$. 
They first extrapolate the $f_B M_B^{1/2}$ linearly to the continuum 
then obtain the decay constant in order to avoid introducing
unnecessary  systematic errors from the heavy meson mass.

\begin{table*}[hbt]
\label{tab:Simulations}
\caption{Decay constant calculation from various groups using
Wilson/clover fermions. The decay constants are in MeV units.
The errors of the MILC result 
are statistical,systematic errors except quenching 
and quenching error. For JLQCD, the errors are statistical, 
systematic errors from chiral exptrapolation and so on, and the
systematics from choosing the scale setting.
The errors of Fermilab group are statistical, systematic errors
from scale setting, and other systematic errors respectively.
The results of MILC, JLQCD and Fermilab are preliminary.}

\begin{tabular*}{\textwidth}{@{}l@{\extracolsep{\fill}}lllllll}
\hline
Group	& MILC \cite{MILC97}  & \multicolumn{2}{c}{JLQCD} \cite{Hashimoto97}
	& \multicolumn{2}{c}{APE} \cite{Allton97}	& UKQCD
\cite{UKQCD94,Wittig97}  & FNAL \cite{Sinead97} \\
Action	& Wilson & Wilson & clover & Wilson & clover   & clover  & clover\\
$\beta$	& 5.72 $\sim$ 6.52 & \multicolumn{2}{c}{5.9,6.1,6.3}
	& \multicolumn{2}{c}{6.0,6.2} & 6.0,6.2 & 5.7,6.1,6.3 \\
$a^{-1}$ from
        & $f_{\pi}$ & \multicolumn{2}{c}{$m_{\rho}$}
        & \multicolumn{2}{c}{$f_{\pi}, m_{\rho},\sigma, f_{K^{\ast}}$ }
	& $m_{\rho}$ 
	& $f_K$ \\
$f_B$    
	& $153(10)({}^{36}_{13})({}^{13}_{0})$
	& \multicolumn{2}{c}{$163(9)(8)(11)$}
	& \multicolumn{2}{c}{$180(32)$} & $160(6)({}^{53}_{19})$
        &$156({}^{13}_{14})({}^{29}_{0})({}^{0}_{20})
         ({}^{0}_{5})({}^{2}_{0})$\\
$f_{B_s}$   
	& $164(9)({}^{47}_{13})({}^{16}_{0})$
	& \multicolumn{2}{c}{$175(9)(9)(13)$}
	& \multicolumn{2}{c}{$205(35)$} & 
        & $177({}^{11}_{12})({}^{39}_{0})({}^{13}_{12})$\\
$f_{D}$   
	& $186(10)({}^{27}_{18})({}^{9}_{0})$
	& \multicolumn{2}{c}{$184(9)(9)(12)$}
	& \multicolumn{2}{c}{$221(17)$} & $185({}^{4}_{3})({}^{42}_{7})$
        & $183({}^{12}_{13})({}^{41}_{0})({}^{9}_{25})$\\
$f_{D_s}$   
	& $199(8)({}^{40}_{11})({}^{10}_{0})$
	& \multicolumn{2}{c}{$203(9)(10)(14)$}
	& \multicolumn{2}{c}{$237(16)$} & 
        & $229({}^{10}_{11})({}^{51}_{0})({}^{3}_{19})$\\
$f_{B_s}/f_B$   
	& $1.10(2)({}^{5}_{3})({}^{3}_{2})$
	& \multicolumn{2}{c}{ }
	& \multicolumn{2}{c}{$1.14(8)$} & $1.22({}^{4}_{3})$
        & $1.17(5)(10)(3)$\\
$f_{D_s}/f_D$   
	& $1.09(2)({}^{5}_{1})({}^{2}_{0})$
	& \multicolumn{2}{c}{ }
	& \multicolumn{2}{c}{$1.07(4)$} &  $1.18({}^{2}_{2})$
        & $1.22(5)(0)(8)$\\
\hline
\end{tabular*}
\end{table*}

\begin{figure}[tb]
\epsfxsize=65mm 
\epsfbox{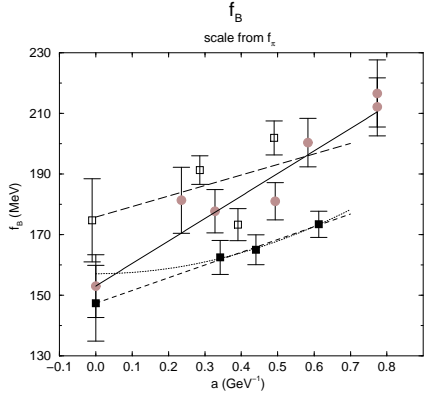 }
\caption{Continuum extrapolation of $f_B$ from MILC Wilson (diamond),
JLQCD Wilson(open square)  and JLQCD clover(filled square)
using $f_\pi$. for the lattice scale.}
\label{fig:decayconst.fpi}
\vspace*{-5mm}
\end{figure}
\begin{figure}[tb]
\epsfxsize=65mm 
\epsfbox{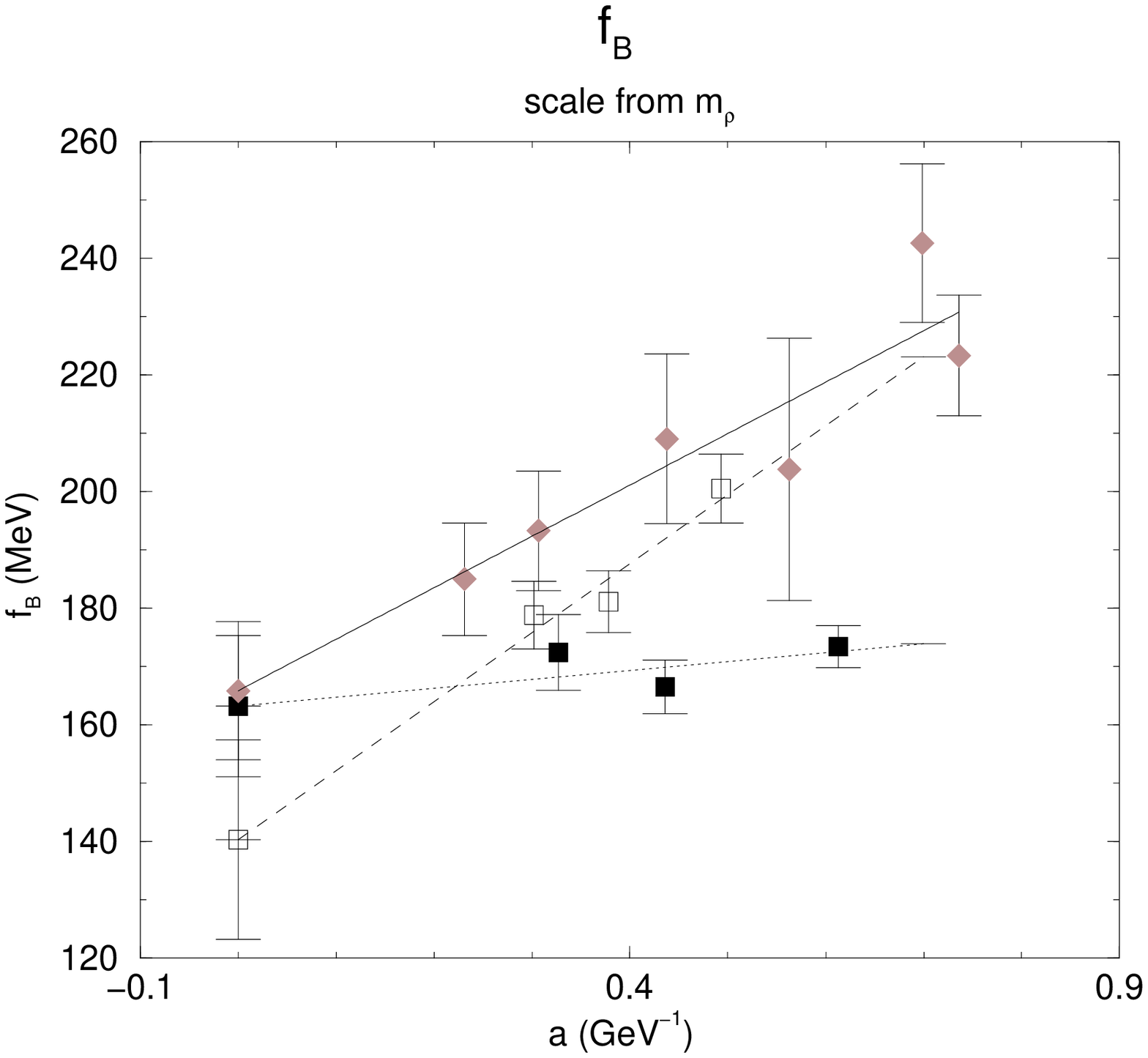 }
\caption{Continuum extrapolation of $f_B$ from MILC Wilson (diamond),
JLQCD Wilson(open square)  and JLQCD clover(filled square)
 using $m_\rho$ for the lattice scale.}
\label{fig:decayconst.mrho}
\vspace*{-5mm}
\end{figure}
\begin{figure}[tb]
\epsfxsize=65mm 
\epsfbox{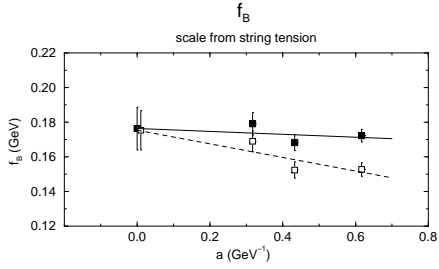}
\vspace*{-20mm}
\caption{Continuum extrapolation of $f_B$.
Filled square and open square are JLQCD Wilson and clover results
 using string tension for the lattice scale.}
\label{fig:decayconst.sigma}
\vspace*{-5mm}
\end{figure}
\begin{figure}[tb]
\epsfxsize=65mm 
\epsfbox{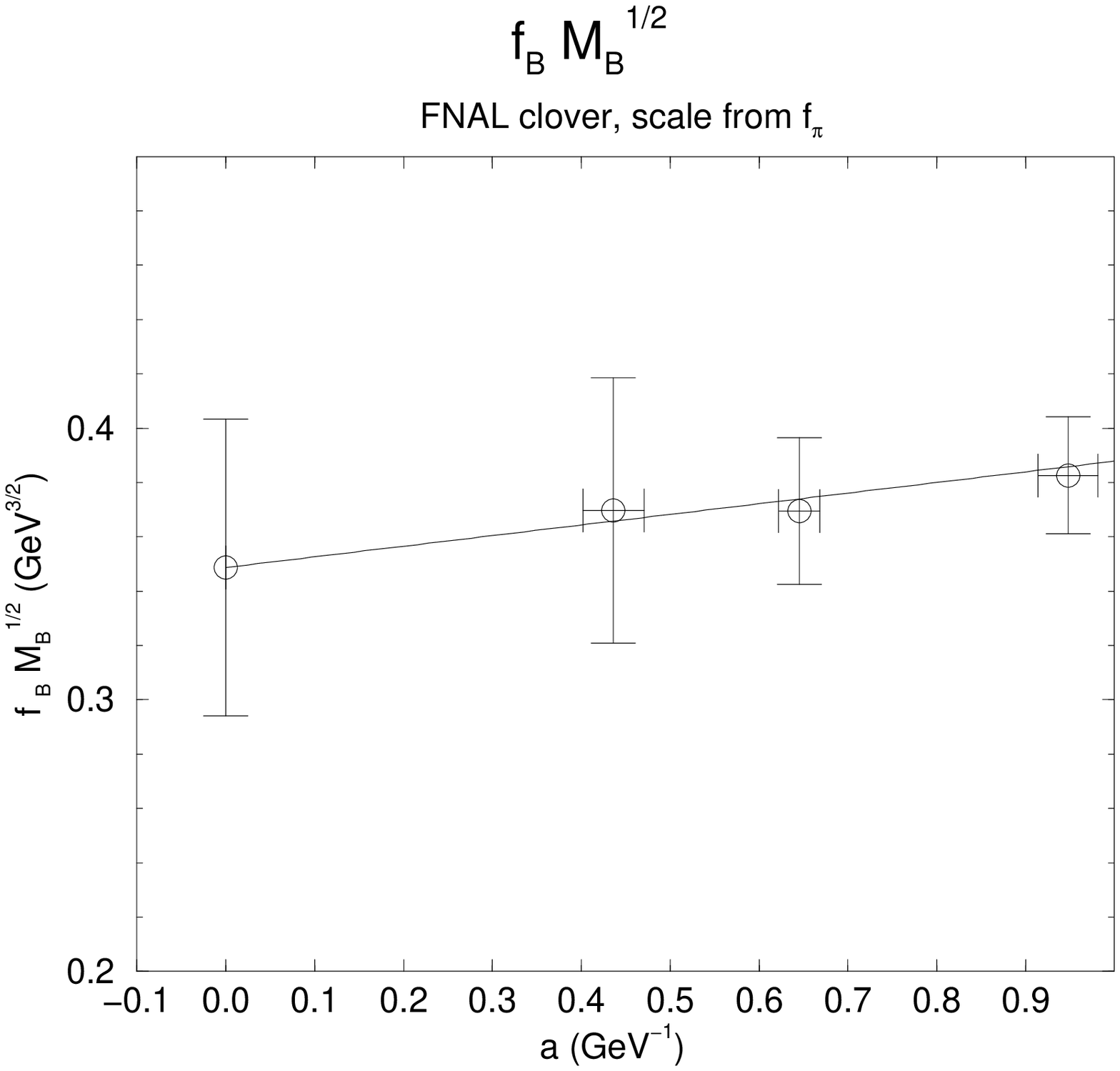}
\caption{Continuum extrapolation of $f_B M_B^{1/2}$ by Fermilab group.
The scale is set from $f_{\pi}$.}
\label{fig:fBsqrtM_FNAL}
\vspace*{-5mm}
\end{figure}

Figures \ref{fig:decayconst.fpi} and \ref{fig:decayconst.mrho}
show the lattice spacing dependences of $f_B$ by JLQCD and MILC collaboration
using Wilson-Wilson and clover-clover fermions, whose scale are
determined from $f_\pi$ and $m_\rho$ respectively.
Figure \ref{fig:decayconst.sigma} shows the data with scales
from the string tension\cite{Balli_Schilling_92}. 
Also, Figure \ref{fig:fBsqrtM_FNAL} shows the lattice spacing
dependence of the $f_B M_B^{1/2}$ by Fermilab Group with scales
set from $f_{\pi}$.

The naive error estimate in the previous section
suggests that the Wilson data should scale linearly in the
lattice spacing $a$ with a rather steep slope.
Taking typical momentum $p$ as $0.3 \sim 0.5GeV$, the data at 
$a=0.5GeV^{-1}$ can typically have 15-25\% $O(a)$ error. 

On the other hand, since the clover data should have $O((ap)^2)$
, $O(\alpha_s ap)$ and $O(\alpha_s^2)$, one expects 5\% errors from 
each contribution, thus the $a$ dependence for the clover data 
should be much smaller than that for Wilson data. 

This is indeed the case for data with scales from $m_\rho$ and string
tension. For example, Figure \ref{fig:decayconst.mrho} shows that 
the clover $f_B$ has very small $a$ dependence, while the Wilson $f_B$ 
has a large $a$ dependence. Moreover, the clover and the Wilson data
seem to have continuum limit which are consistent with each other.
It seems that the MILC Wilson data are slightly higher than JLQCD
Wilson data by a few \%, which will be discussed later.

Now the situation for $f_B$ with scales from $f_\pi$ is not clear.
The linear slope of Wilson $f_B$ is entirely different between JLQCD
and MILC collaboration. Also within JLQCD data the continuum limits
from Wilson and Clover are not in agreement within error.

Here, three question arises:

\begin{enumerate}
\item Are the renormalization factors $Z_A$ consistent between two 
groups?
\item Are the raw data consistent?
\item Are there any problems in setting the scale, especially 
from $f_\pi$?
\end{enumerate}

In the following, we will consider these problems in detail.  
 
\subsubsection{Renormalization factor}
\begin{figure}[tb]
\epsfxsize=7.5cm
\epsfbox{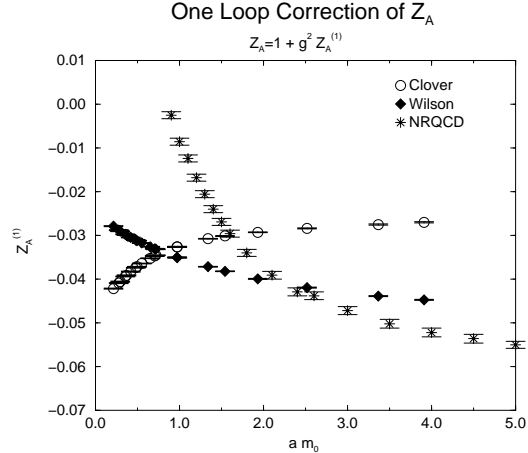}
\caption{One-loop coefficient of $Z_A$ for Wilson
   (circles), clover (diamonds) and NRQCD (stars) as a
   function of the bare heavy quark mass.}
\label{fig:Z_A}
\vspace*{-5mm}
\end{figure}

As was mentioned in the introduction, one-loop renormalization factor
$Z_A$ for the heavy-light axial vector current 
has been calulclated as a function of the heavy quark
mass\cite{Ishikawa97,Aoki_etal_97}. 
Figure\ref{fig:Z_A} shows the mass dependence of the
one-loop coefficient of $Z_A$.
It is found that although there is indeed heavy
quark mass dependences in $Z_A$, for $a m_0 \sim 1-3$ , the difference
between the $Z_A$ for heavy-light axial current and that for the
massless light-light axial current is around 2-3 percent for 
$g^2 \sim 2-3$. MILC collaboration uses $Z_A$ in massless 
limit even for the heavy-light current, while JLQCD uses 
slightly smaller $Z_A$ with mass dependence included. 
This explains part of the descrepancy between JLQCD and MILC.

It should also be noted that the scale $q^{\ast}$ of the
coupling $\alpha_v$ differs between JLQCD and MILC,  
which are $q^{\ast}=1/a$ and
$q^{\ast}=2.32/a$\cite{Bernard_Golterman_97} respectively.
Using the same $q^{\ast}$ makes the results come closer
again by a few percent. 
\subsubsection{Comparison of the raw data}
\begin{figure}[tb]
\epsfxsize=7.5cm
\epsfbox{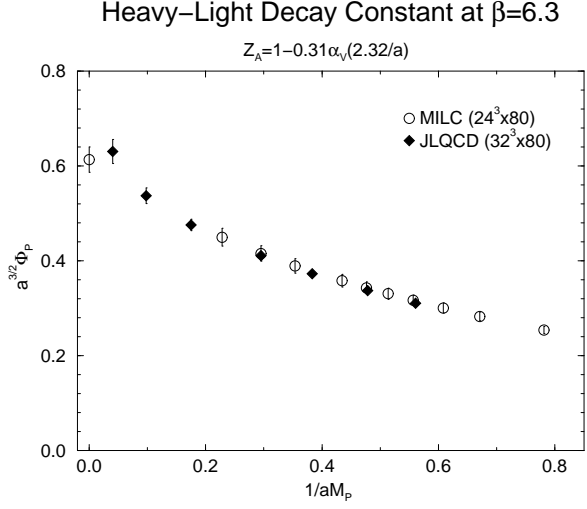}
\vspace*{-40mm}
\caption{Direct comparison of JLQCD and MILC.}
\label{fig:MILC_vs_JLQCD}
\vspace*{-5mm}
\end{figure}

We compare raw data of $f_P\sqrt{M_P}$ from  MILC and JLQCD
Wilson.  
Since MILC and JLQCD
has a slightly different analyses, it is not easy to see whether 
the two groups are consistent with each other from their quoted values. 
In Figure \ref{fig:MILC_vs_JLQCD}, we give a plot of
$f_P\sqrt{M_P}$ at $\beta=6.3$ where the data are available
from both groups.
The JLQCD's raw data are re-analyzed 
in the same way as MILC so that they have the same renormalization
factor $Z_A$ with $q^{\ast}=2.32/a$.

We find that two results are in perfect agreement with each other.

\subsubsection{Scale setting}
The last issue is the scale setting.
Since accuracy required in the lattice calculation of $f_B$ is around 
10\%, the scale $a^{-1}$ used by various groups are in fact
important. Of course, there should be
differences between the scales $a^{-1}$ from different inputs, which 
are nothing but the queching effects. However, any disagreement in
the scales $a^{-1}$ from the same inputs due to the fitting procedure
etc. can be the source of systematic error which has to be removed for
the high precision calculation of $f_B$. Another problem is the 
consistency of the results in the continuum limits from different
actions. Although the scale from the same 
input can differ for Wilson and clover for finite cutoff, the
continuum limit should agree. If they differ,
there should be some systematic error such as fitting procedure, 
renormalization, or the way to take the continuum limit. 

 \begin{itemize}
 \item $a^{-1}$ from $m_\rho$ \\
   Figure \ref{fig:mrho_st_all} shows the ratio of the scale $a^{-1}$
   from $m_\rho$ to that from the string tension
   (obtained by Balli-Schilling\cite{Balli_Schilling_92}) for
   Wilson and clover fermions.  
 
   It can be seen that the $a^{-1}(m_\rho)$ from JLQCD, CPPAX, MILC
   and  APE  Wilson data agree with each other over the whole range. 
   The clover JLQCD, APE and UKQCD data have consistenly smaller
   values but the  continuum limit agrees with that for Wilson 
   within statistical errors.
 
   \begin{figure}[tb]
     \epsfxsize=7.5cm
     \epsfbox{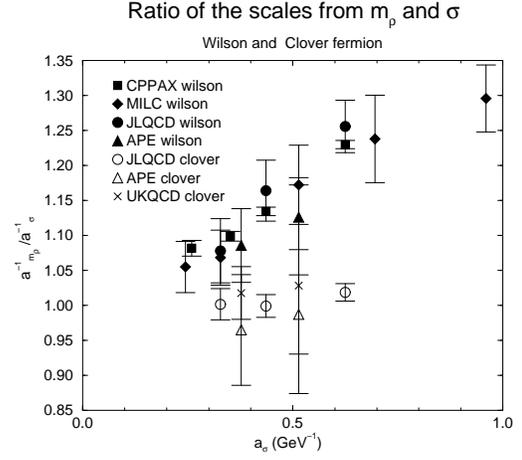}
     \caption{$a^{-1}$ from $m_\rho$. Filled circle is the JLQCD
       Wilson data, filled diamond is the MILC Wilson data,
       ,square is the CPPAX Wilson data, Filled triangle is the APE 
	Wilson data. Open circles are JLQCD clover data, open
	triangles are APE clover and crosses are UKQCD clover data.
        } 
     \label{fig:mrho_st_all}
     \vspace*{-5mm}
   \end{figure}
 
 \item $a^{-1}$ from $f_\pi$ \\
   Figure \ref{fig:fpi_st_all} shows the ratio of the scale $a^{-1}$
   from $f_\pi$ to that from the string tension
   for Wilson and clover fermions. 
 
   The result from JLQCD Wilson and CP-PACS Wilson are in agreement,
   whereas MILC results with linear chiral extrapolation are more than
   10\%  smaller than those by JLQCD and CP-PACS over almost the whole range. 
   MILC results with quadratic chiral extrapolation lie in between. 
   It should be noted that the perturbative renormalization scale 
   $q^{\ast}$ used for $\alpha_v$ differs between JLQCD-CP-PACS and MILC, 
   which are $q^{\ast}=1/a$ and $q^{\ast}=2.32/a$ respectively.
   Using the same $q^{\ast}$ makes the results come closer, but by only a 
   few percent. The origin for the rest descrepancy is not clear: it may
   either come from the uncertainty of the linear extrapolation or some 
   other effects such as finite volume effects.
   APE wilson and UKQCD clover data have large $a$dependences, 
   while APE clover data are consistent with JLQCD data.
   \begin{figure}[tb]
     \epsfxsize=7.5cm
     \epsfbox{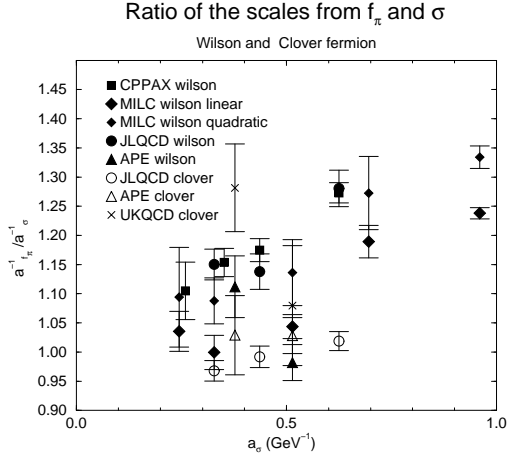}
     \caption{$a^{-1}$ from $f_\pi$. Filled circle is the JLQCD
       Wilson data, filled large(small) diamonds are the MILC Wilson
	data with linear(quadratic) chiral extrapolation
       ,square is the CPPAX Wilson data, Filled triangle is the APE 
	Wilson data. Open circles are JLQCD clover data, open
	triangles are APE clover and crosses are UKQCD clover data.
        } 
     \label{fig:fpi_st_all}
     \vspace*{-5mm}
   \end{figure}

\end{itemize}

Therefore, using Wilson fermions 
JLQCD and MILC collaborations obtains
consistent scale determined from $m_{\rho}$, which seem to have 
consistent continuum limit with that by JLQCD using clover fermions.

For the scale determined from $f_{\pi}$, the results with Wilson
fermions from JLQCD and MILC collaborations do not agree.
The continuum limit of the clover results from JLQCD are 
consistent with that by MILC Wilson data but not with JLQCD Wilson
data.

\subsubsection{Effects of Quenching}
\begin{figure}[tb]
\epsfxsize=7.5cm
\epsfbox{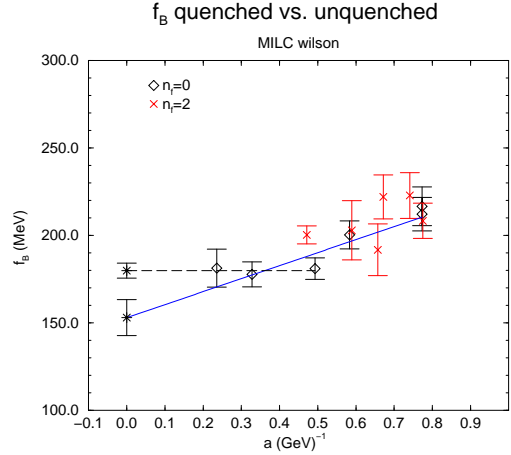}
\caption{$f_B$ for quenched and unquenched Wilson by MILC collaboration. }
\label{fig:fb_que_vs_dyn}
\vspace*{-5mm}
\end{figure}

\begin{figure}[tb]
\epsfxsize=7.5cm
\epsfbox{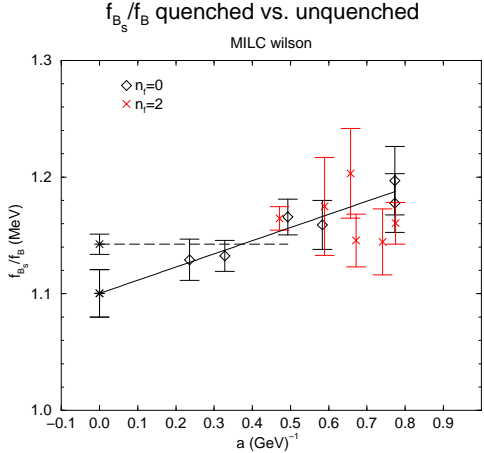}
\caption{$f_{B_s}/f_B$ for quenched and unquenched Wilson by MILC 
collaboration . }
\label{fig:fbs_over_fb_que_vs_dyn}
\vspace*{-5mm}
\end{figure}

Let us now consider the systematic error from the quenched
approximation. MILC collarboration estimated effects of quenching in
two ways: (1) by comparing the 
a smallest dynamical lattice, with the quenched 
results interpolated to the same lattice spacing and 
(2) redoing the calculation switching from the $f_{\pi}$ scale to the
$m_{\rho}$  scale. They take the biggest deviation from the central value 
as the error estimate. 
In Figs.\ref{fig:fb_que_vs_dyn},\ref{fig:fbs_over_fb_que_vs_dyn}
quenched results and unquenched results of $f_B$ and $f_{B_s}/f_B$ 
are compared. 

\subsubsection{Summary of $f_B$}
The study of the previous section shows that Wilson data 
are essentially in agreement. What makes the result different 
is the choice of the renormalization factor and the scale.
We observe that $m_{\rho}$ is a reliable quantity to set the scale,
while there are still uncertainties for $f_{\pi}$. 
The final result of 
MILC Wilson, JLQCD Wilson/clover, UKQCD clover, FNAL clover
for $f_B$ are in good agreements. The central value of APE result is higher 
than others  but within errors. 
The global value of $f_B$ in the quenched approximation 
I quote is JLQCD result is: 
\begin{eqnarray}
f_B & = &163(9)(8)(11) {\rm MeV}.
\end{eqnarray}
For the quenching error, I quote the error estimated by MILC 
collaboration which is slightly less than 10 \%.
\subsection{Bag Parameters}
The Bag parameter is calculated using static heavy and clover 
light by UKQCD for $\beta=6.2$\cite{Ewing96}, by Gimenez and 
Martinelli~\cite{GM} for
$\beta=6.0$ in the quenched approximation, and by MILC~\cite{MILC_BB_97} for
$\beta=5.85$ in the quenched  approximation, and for $\beta=5.5,5.6$
with $n_f=2$. 

\begin{table}[tbp]
\setlength{\tabcolsep}{0.3pc}
\begin{center}
\caption{Results for the Bag parameters. The results from 
MILC collaboration are preliminary.}
\begin{tabular}{llll}
\hline
Group & $\beta$ & $B_B(m_b)$ & $\hat{B}_B^{NLO}$ \\ 
\hline
MILC\cite{MILC_BB_97}
      & 5.85($n_f=0$) & 0.94(4)   &  		\\ 
      & 5.5($n_f=2$)  & 0.87(2)   &  $\sim$ 1.4	\\ 
      & 5.6($n_f=2$)  & 0.94(3)   &  		\\ 
 \hline
G+M \cite{GM}
      & 6.0	      & 0.63(4)   & 1.00(6)     \\ 
      & 	      & 0.73(4)   & 1.16(6)     \\ 
UKQCD\cite{Ewing96}
      & 6.2	      & 0.69(4)   & 1.10(${}^5_6$)(${}^3_2$) 
						\\ 
\hline
\end{tabular}
\label{tab:result_BB}
\end{center}
\end{table}

\subsection{Semileptonic Form Factors}
In spite of many years of effort~\cite{Flynn96} 
to compute the semileptonic form
factors, heavy-light to light-light such as 
$ B \rightarrow \pi,\rho$ is still a big challenge to 
lattice QCD. The reason for the difficulty is that the form factors
at lower $q^2$ has large statistical errors since the noise of the 
three point functions grows as 
$ \sim \exp( E_{\pi,\rho} t )$. 
Moreover, the systematic error due to the finite lattice spacing 
which is of $O(ap)$ for Wilson fermion and the $O((ap)^2)$ and $O(\alpha a)$
for clover fermion makes it hard to obtain the correct $q^2$
dependence. 
This is different from the situation in $D$ meson 
decay~\cite{Debbio97,Abada94,Allton95,BKS92,Wu97,LMMS92,BKS92,BG95}, 
in which case the highest momentum of the final state meson is less than 
1 GeV so that the lattice simulation can cover the whole 
kinematically allowed range without encountering the above
difficulties.

\begin{table*}[hbt]
\label{tab:Form}
\caption{Simulations of heavy-light to light-light semileptonic decay
form factors by various groups.
The results from JLQCD collaboration and Hiroshima collaboration 
are preliminary.}
\begin{tabular*}{\textwidth}{llllllllll}
\hline
Group			& $\beta$		& Heavy(Light) 	Actions	
			& $m_Q$	 		& $m_q$ \\
JLQCD\cite{Tominaga97}	& 5.9			& clover(clover)	
			& $m_c \sim m_b$ 	& chiral limit	\\
Hiroshima
\cite{Matsufuru97}	& 5.8			& NRQCD(Wilson)		
			& $m_c \sim m_b$	& chiral limit	\\
Wuppertal\cite{Wu97}	& 6.3			& Wilson(Wilson)	
			& $\sim m_c$		& chiral limit	\\
UKQCD\cite{Bowler95,UKQCD}	
			& 6.2			& clover(clover)	
			& $\sim m_c$		& $\sim m_s$  \\
FNAL\cite{Simone95}	& 5.9			& clover(clover)	
			& $m_c,m_b,\infty$	& $\sim m_s$ \\
APE\cite{Allton95}	& 6.0			& clover(clover)	
			& $\sim m_c$		& chiral limit	\\
LANL\cite{BG95}		& 6.0			& Wilson(Wilson)	
			& $\sim m_c$		& chiral limit	\\
ELC\cite{Abada94}	& 6.4			& Wilson(Wilson)	
			& $\sim m_c$		& chiral limit	\\
\hline
\end{tabular*}
\end{table*}

Table~\ref{tab:Form} shows the list of the simulations on 
heavy-light to light-light form factors.
In Earlier works simulations with heavy quark in the charm quark mass range
were done mostly to study the $D$ meson decay. However, the results can be 
extrapolated in $1/M$ to the $B$ meson decay.
There are two approaches to extrapolate the results, 
both of which uses $1/M$ scaling from the charm quark mass range for
the heavy quark mass. The first approach, taken by UKQCD, for example, 
 is to scale the form factors near $q^2_{max}$ 
assuming the following scaling laws predicted from the
heavy quark effective theory:
\begin{eqnarray}
f^0, A_1, T_2 & \sim & M^{-1/2} \nonumber \\
f^+,V,A_2,T_1 & \sim & M^{1/2} \nonumber
\end{eqnarray}

In this approach, the range of $q^2$ is limited to the range 
near $q^2_{max}$. In order to know the form factors 
for smaller $q^2$, one has to extrapolate the result 
by fitting the form factors to some model such as the pole model. 

Another approach, taken by the Wuppertal group, for example, 
is to obtain the form factors at $q^2=0$ 
for charm quark mass range, which can reliably be done by 
interpolation. The form factors are then scaled in $1/M$.
The second method requires the knowledge of the 
scaling law at $q^2=0$, which heavy quark effective theory 
cannot predict.

If one assumes the pole model or its generalization, these 
two approaches are related to each other since
the form factor ansatz in the model relates the 
$1/M$ scaling behavior near $q^2_{max}$ and at $q^2=0$.

Fermilab group~\cite{Simone95} has studied the $1/M$ dependence of 
the form factors $f^+,f^0$ with heavy quark mass 
at the static, the bottom and the charm quark mass and with light
quark mass at the strange quark.
Recently, JLQCD collaboration\cite{Tominaga97} and Hiroshima collaboration\cite{Matsufuru97} performed 
calculations of the semileptonic form factors 
near $q^2_{max}$ using clover heavy - clover light
and NRQCD heavy - Wilson light. Both groups treat the heavy 
quark action as the nonrelavisitic effective action, hence 
the simulations at the b quark mass range are possible.

In the following, we discuss the scaling behavior in $1/M$.

\subsubsection{$1/M$ scaling near $q^2_{max}$}

\begin{figure}[tb]
\epsfxsize=7.5cm
\epsfbox{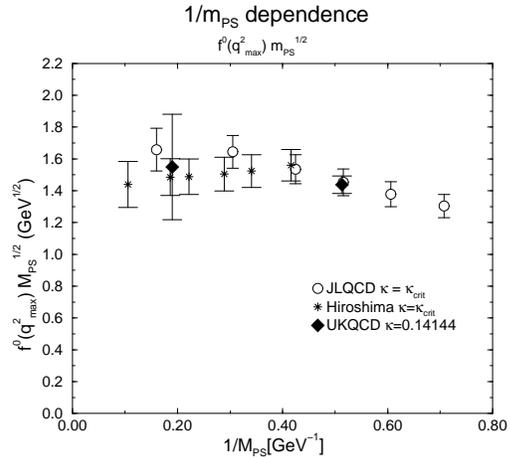}
\caption{$f^0(q^2_{max}) M^{1/2}$ from various groups}
\label{fig:f0}
\vspace*{-5mm}
\end{figure}

Fig.\ref{fig:f0} shows  $M$ dependence of $f^0(q^2_{max})M^{1/2}$.
It is found that the $1/M$ scaling behavior is consistent 
with the prediction of the heavy quark effective theory
 and that the $1/M$ correction 
is small. JLQCD and Hiroshima gives consistent result. 
They are also consistent with the result of UKQCD which is 
obtained by scaling the form factor near the charm mass range
up to the bottom quark mass.

\begin{figure}[tb]
  \epsfxsize=7.5cm
  \epsfbox{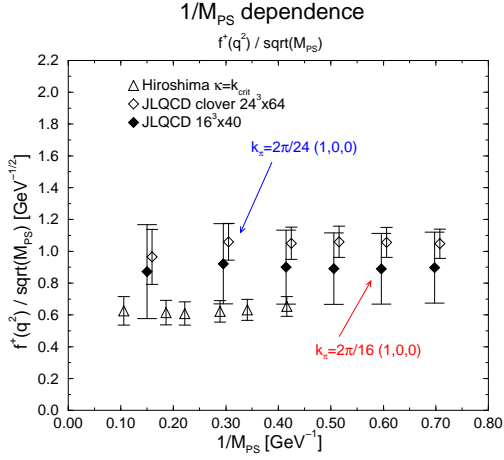}
  \caption{$f^+(q^2)/\protect\sqrt{M_{PS}}$ in the chiral
    limit as a function of $1/M_{PS}$. 
    Open diamonds correspond to the results on a $16^3 \times
    40$ lattice and filled diamonds on a $24^3 \times 64$ lattice. 
    The spatial momentum of $B$ meson is zero and the pion 
    is moved with the smallest momentum in lattice units.
    Trianlges are the results from Hiroshima NRQCD + Wilson 
    results}
 \label{fig:fp}
\vspace*{-5mm}
\end{figure}

Fig.\ref{fig:fp} is the plot of the $M$ dependence of
$f^+(q^2)/M^{1/2}$ with the pion momemtum $k_{\pi}=\frac{2 \pi}{L}\cdot
(1,0,0)$. Again the data confirms the $1/M$ scaling behavior predicted by 
the heavy quark effective theory. 

\begin{figure}[tb]
  \epsfxsize=7.5cm
  \epsfbox{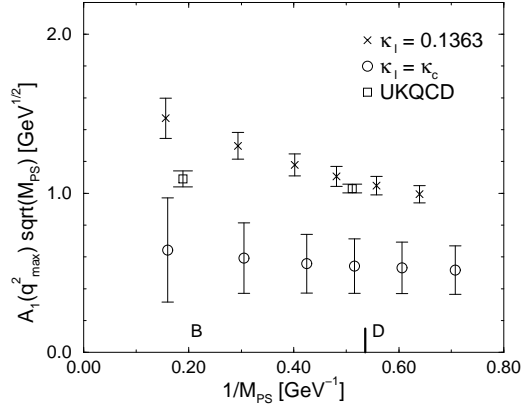}
  \caption{Squares are the data of $A_1(q^2_{max}) \protect\sqrt{M_{PS}}$
  from UKQCD  which is extrapolated  $1/M$  with the light quark at 
  $k_l=0.14144$, and crosses and circles  are those of
  JLQCD on a $16^3\times40$ lattice at $\kappa=0.1363$ and  $\kappa_c$ 
  as a function of  $1/M_{PS}$. }
  \label{fig:A1}
\vspace*{-5mm}
\end{figure}

In Fig.\ref{fig:A1} $A_1(q^2_{max})M^{1/2}$ is plotted against $1/M$. 
The $1/M$ scaling behavior is again consistent 
with the heavy quark effective theory.
There is a significant discrepancy of the JLQCD data with the UKQCD data.
JLQCD group found that the $A_1(q^2_{max})$ decreases
rather rapidly as the light quark mass approaches to the chiral limit.
Since the result of UKQCD is obtained not in the chiral limit 
but with the strange light quark, the descrepancy may be explained 
by the difference in the light quark mass.

\begin{figure}[tb]
\epsfxsize=7.5cm
\epsfbox{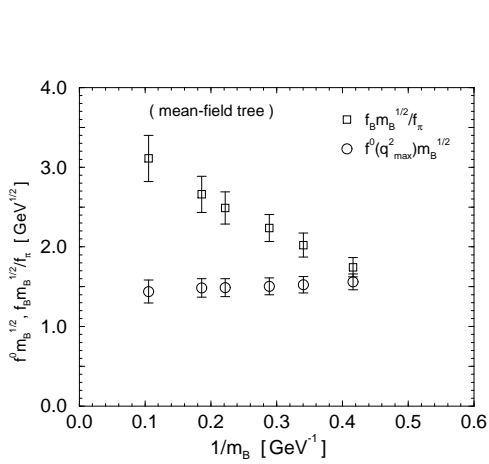}
\caption{Comparison of $f^0(q^2_{\,max})$ with $f_B/f_{\pi}$ 
at mean-field tree level. }
\label{fig:fBfpi}
\vspace{-5mm}
\end{figure}

There is another problem in the chiral limit.
Soft pion theorem predicts the relation $f^0(q^2_{max}) =
f_B/f_{\pi}$\cite{BD92,KK94,Soft_Pion_Theorem}. Fig.\ref{fig:fBfpi} gives plots of $f_B M^{1/2}/f_{\pi}$ and
and $f^0(q^2_{max})$ by Hiroshima Group. 
The two quantity have a large descrepancy in the heavy quark 
mass limit. JLQCD group also finds similar phenomena. 
Possible reasons are the breaking of chiral symmetry on the 
lattice, the renormalization, the systematic uncertainty 
of the chiral extrapolation, or the breakdown of the soft pion theorem
itself in the semileptonic decay. This needs further study.

\begin{figure}[tb]
\epsfxsize=7.5cm
\epsfbox{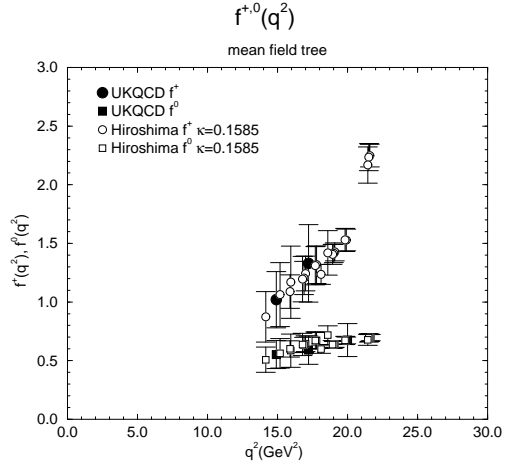}
\caption{$q^2$ dependence of form factors.}
\label{fig:f+0_all}
\vspace*{-5mm}
\end{figure}

The $q^2$ dependences of the form factors $f^0,f^+$ for $B$ meson with
strange light quark are shown in Fig.\ref{fig:f+0_all}. 
The Hiroshima Groups result and the UKQCD result extrapolated from 
the charm quark mass range are consistent. 
In the chiral limit, the statsitical error of the form factors with
nonzero recoil becomes very large so that the pion signal with 
only zero and the smallest unit of momentum can be observed.
Clearly to obtain the form factors in the chiral limit with pion
momentum of 
$1 GeV$, one needs statistics of a few hundred configurations.

For strange light quark mass, the statistics is good enough 
to extract $q^2$ dependence upto $k_\pi \sim 1 GeV$,
in fact $B \rightarrow \pi + l + \overline{\nu}$
semileptonic decay rate clearly exhibits the pion spectrum,
however, the present level of the systematic error in the $q^2$ dependence is 
$O(ap)$ for Wilson fermion which gives a large uncertainty in the decay rate.
Therefore the use of improved actions and calculations on several 
lattices with larger $\beta$ are required as was the case for the 
heavy-light decay
constant. 
 
\subsubsection{$1/M$ scaling at $q^2=0$}
According to the prediction of the light cone sum rule (LCSR)\cite{Ball},
all the form factors at $q^2=0$ scales as $M^{-3/2}$.
Using pole-type models with the $1/M$ scaling behavior which are 
consistent both with HQET and LCSR, UKQCD group fitted their lattice
data to the form factors for $B \rightarrow \pi,\rho$
to obtain the form factors at $q^2$.

On the other hand, Wuppertal group scaled the form factors
at $q^2=0$ obtained from simulation at the charm quark range
without necessarily assuming LCSR prediction. Instead, they tried three
different ans$\ddot{a}$tze for the $1/M$ scaling, which are
$F(0) \sim M^{-1}, M^{-3/2} ~\rm{and}~ M^{-1/2}$. The present
statistics does not distinguish the three different fits so that
they get consistent results.
The situation is not yet clear with the present statistics.
Further study is need to test the scaling behavior at $q^2$.
The results for the semileptonic form factors at $q^2=0$ 
are listed in Table~\ref{tab:Form_q2_0}.

\begin{table*}[hbt]
\label{tab:Form_q2_0}
\caption{Semileptonic decay form factors at $q^2=0$.}
\begin{tabular*}{\textwidth}{rrrrr}
\hline
\multicolumn{5}{c}{$B\rightarrow \pi/\rho$} \\ 
\hline
Group		&$f(0)$		& $V(0)$		& $A_1(0)$
		& $A_2(0)$      \\
Wuppertal\cite{Wu97}
		& $0.43(19)$	& $0.65(15)$ 		& $0.28(3)$
		& $0.46(23)$	\\
UKQCD\cite{Debbio97}
		& $0.27(11)$	& $0.35({}^{6}_{5})$	&$0.27({}^{5}_{4})$
		& $0.26({}^{5}_{3})$
				 \\
APE a\cite{Allton95}
		& $0.29(6)$	& $0.45(22)$		& $0.29(16)$
		& $0.24(56)$	 \\
APE b\cite{Allton95}
		& $0.35(8)$	& $0.53(31)$		& $0.24(12)$
		& $0.27(80)$	\\
ELC a\cite{Abada94}
		& $0.26(12)(4)$	& $0.34(10)$		& $0.25(12)$
		& $0.25(6)$	\\
ELC b\cite{Abada94}
		& $0.30(14)(5)$	& $0.37(11)$		& $0.27(5)$
		& $0.49(21)(5)$	\\
\hline
\multicolumn{5}{c}{$D\rightarrow K/K^{\ast}$ }\\ 
\hline
Group		&$f(0)$		& $V(0)$		& $A_1(0)$
		& $A_2(0)$	 \\
Wuppertal\cite{Wu97}
		& $0.78(5)$	& $1.27(16)$ 		& $0.67(4)$
		& $0.67(13)$	\\
LMMS\cite{LMMS92}
		& $0.63(8)$	& $0.86(10)$ 		& $0.53(3)$
		& $0.19(21)$	\\
BKS\cite{BKS92}	& $0.90(8)(21)$	& $1.43(45)(49)$	& $0.83(14)(28)$
		& $0.59(14)(24)$ \\
APE\cite{Allton95}
 		& $0.77(4)$	& $1.16(16)$		& $0.61(5)$
		& $0.49(34)$	\\
ELC\cite{Abada94}
		& $0.60(15)(7)$	& $0.86(24)$		& $0.64(16)$
		& $0.40(28)(4)$	\\
LANL\cite{BG95}	& $0.73(5)$	& $1.27(8)$		& $0.66(3)$
		& $0.44(16)$	\\
UKQCD\cite{Bowler95}
		& $0.67({}^{7}_{8})$	
				& $1.01({}^{30}_{13})$	& $0.70({}^{7}_{10})$
		& $0.66({}^{10}_{15})$	
				 \\
\hline
\end{tabular*}
\end{table*}

\section{Summary}
In summary, the heavy-light decay constant in the quenched
approximation now has a precision of less than 20\% 
once one uses the appropriate scale setting. 
For higher precision of less than 10\%, full one loop 
renormalization including the operator mixing will be important.
Comparison with decay constants with NRQCD+clover will be 
another consistency check of the result. 

A lot of more work with higher statistics as well as use of improved
action are needed for the semileptonic form factors. 
Reducing the systematic errors are especially important to 
understand the $q^2$ dependence and the puzzling violation 
of the soft pion theorem. 
Studies of $q^2\sim q^2_{max}$ are important, because
the future experiments such as KEK B factory which will start in 1999
are expected to have the luminosity of about $10^{34}cm^{-2} s^{-1}$ 
produing $10^8$ $B$ $\overline{B}$ pairs after 3-4 years of running.
which is 30 times more than the CLEO data. One may expect to see
enough experimental data even for the small recoil region.

Studies at the scaling behavior at $q^2=0$ around charm quark
mass range to test the LCSR with higher statistics are also very important. 

\noindent{\bf Acknowledgement}
I would like to thank J. Flynn, T. Yoshie, A. Ali Khan,
S.G$\ddot{u}$sken, K.Schilling 
S. Ryan, J.Simone, A. El-Khadra, 
L. Giusti, H. Wittig, C. Bernard, S.Gottlieb, C. McNeile, 
for providing me with information on their results and discussions.  
I am also grateful to  S. Hashimoto, S. Tominaga,
H. Matufuru, K-I. Ishikawa, N. Yamada and other members of JLQCD
collaboration for fruitful discussions.

This work was partly supported by Monbusho International
Scientific Reserach Program (No. 08044089).
 


\begin{thebibliography}{99}
 \bibitem{El-Khadra_Kronfeld_Mackenzie_97}
   A.X. El-Khadra, A.S. Kronfeld, P.B. Mackenzie,
   Phys. Rev. {\bf D55} (1997) 3933.
\bibitem{Review_NRQCD_97}
 A. Ali-Khan, in these procedings.
\bibitem{Luscher_Weisz_96}
  M. L\"uscher and P. Weisz, 
  Nucl. Phys. \textbf{B479} (1996) 429.
\bibitem{Ishikawa97} 
  K-I. Ishikawa et al., these proceeding.
\bibitem{JSHEP} 
  J.~Shigemitsu, hep-lat/9705017.
\bibitem{Bernard_Golterman_97}
  C. Bernard, M Golterman and C. McNeile  in preparation. 
\bibitem{Allton97}
C.R. Allton et al. hep-lat/9703002. 
\bibitem{Henty95} D.S.Henty et al. 
Phys. Rev. \textbf{D51}(1995) 5323
\bibitem{Martinelli93} G. Martinelli et al. 
Phys. Lett. B311 (1993) 241; 
Erratum Phys. Lett. B317(1993)660.
 \bibitem{UKQCD94}
   R.M.Baxter et al. Phys. Rev. \textbf{D49} (1994) 1594,
 \bibitem{Ewing96}
   A.K.Ewing et al. Phys. Rev. \textbf{D56} (1996) 3526 ,
\bibitem{Balli_Schilling_92}G. S. Balli and K. Schilling, 
  Phys. Rev. \textbf{D46} (1992) 2636.
\bibitem{MILC97}
  MILC  Collaboration in these proceedings.
\bibitem{Hashimoto97}
  JLQCD Collaboration (S.~Aoki {\it et al.}), presented by
  S. Hashimoto, in these proceedings. 
\bibitem{Bowler95}	
  K. Bowler {\it et al.}  Phys. Rev. \textbf{D51} (1995) 4905.
\bibitem{Wittig97}
H. Wittig hep-lat/9705034.
\bibitem{Sinead97}
Fermilab group, 
(presented by S. Ryan) these proceedings; 
FNAL Pub-97/322-T.
\bibitem{Aoki_etal_97}
  S. Aoki, S. Hashimoto, K-I. Ishikawa and T. Onogi,
  in preparation. 
\bibitem{GM}
 V. Gimenez and G.Martinelli, Phys. Lett. B 347 (1997) 135
\bibitem{MILC_BB_97}
 MILC Collaboration, these proceedings.
\bibitem{Flynn96}
  For a review, see J. Flynn, Nucl. Phys. B (Proc. Suppl.) {\bf 53} (1997) 168.
 \bibitem{Debbio97}
   L.D.Debbio {\it et al.}, CPT-97/P.3505, SHEP-97/13
 \bibitem{Allton95}
   C.R. Allton {\it et al.}, Phys. Lett. {\bf B345} (1995) 513,
 \bibitem{Abada94}
   Abada {\it et al.}, Nucl. Phys. {\bf B416} (1994) 675.
\bibitem{Wu97}
Wuppertal group, 
(presented by S. G$\ddot{u}$sken) these proceedings.
 \bibitem{LMMS92}
   V. Lubicz {\it et al.}, Phys. Lett.B 345 (1992) 415.
 \bibitem{BKS92}
   C. Bernard {\it et al.}, Phys. Rev.	\textbf{D43} (1992) 2140;
\bibitem{BG95}	
  T. Bhattacharya and R. Gupta Nucl. Phys.B (Proc.Suppl.) 42 (1995) 935.
\bibitem{Tominaga97} 
   JLQCD Collaboration ( presented by S. Tominaga ), 
   these proceedings.
\bibitem{Matsufuru97}
S. Hashimoto {\it et al.} (presented by H. Matsufuru), 
these proceedings.
 \bibitem{UKQCD}
   D. R. Burford {\it et al.}, Nucl. Phys. {\bf B447} (1995) 425,
   J. M. Flynn {\it et al.}, Nucl. Phys. {\bf B461} (1996) 327.
 \bibitem{Simone95}
   J. Simone, Nucl. Phys. B (Proc. Suppl.) {\bf 47} (1997) 17.
\bibitem{BD92} 
   G. Burdman and J. F. Donoghue, Phys. Lett. B280 (1992) 287,
   M. B. Wise, Phys. Rev. D45 (1992) 2188
\bibitem{KK94} 
   N. Kitazawa and T. Kurimoto, Phys. Lett. B323 (1994) 65.
\bibitem{Soft_Pion_Theorem}
  See, {\it e.g., }, G. Burdman {\it et al.,}
  Phys. Rev. {\bf D49} (1994) 2331.
\bibitem{Ball}
 P. Ball, hep-ph/9605233.
\end{thebibliography}
\end{document}